\documentclass[%
reprint,
 amsmath,amssymb, aps
]{revtex4-2}
\usepackage{xcolor}

\usepackage{graphicx}% Include figure files
\usepackage{dcolumn}% Align table columns on decimal point
\usepackage{bm}% bold math
\usepackage{multirow}% bold math
\usepackage{caption}

\usepackage{natbib}
\setcitestyle{authoryear,open={(},close={)}}

\usepackage{blindtext}
\usepackage[]{xcolor}

\begin{document}

%\title{{A general prescription to understand the electron capture mechanisms during ion-atom collisions}}
\title{{Study of bremsstrahlung radiation coming from a transmission type x-ray generator and its application on EDXRF technique}}
%\title{{Comprehending the complex M x-ray emission mechanisms in heavy ion-atom collisions}}

\author{Soumya Chatterjee$^1$, Sumana Ghosh$^1$ and D. Mitra$^1$}
\affiliation{$^1$Department of Physics, University of Kalyani, Kalyani, West Bengal-741235, India.}
%C.C. Montanari$^2$\affiliation{$^2$ Instituto de Astronomía y Física del Espacio, CONICET and Universidad de Buenos Aires, Buenos Aires, Argentina}
%\affiliation{$^{2}$Inter-University Accelerator Centre, Aruna Asaf Ali Marg, Near Vasant Kunj, New Delhi-110067, India.}
%\thanks {Email:\hspace{0.0cm} %nanditapan@gmail.com. Present address: 1003 Regal, Mapsko Royal Ville, Sector-82, Gurgaon-122004, India.
\begin{abstract}
{We have demonstrated the use of the bremsstrahlung radiation from a small, portable, transmission type x-ray generator that can be used in energy dispersive x-ray fluorescence technique as a tool to perform non-destructive elemental analysis of solid samples employed in inter-disciplinary science research. %The bremsstrahlung radiation \rrt{can be taken} from a small, portable, transmission type x-ray generator. 
As our knowledge goes, till date, this is the first attempt to generate the bremsstrahlung spectrum theoretically from a portable transmission type x-ray generator and compared it with the actual observation.  Theoretically generated bremsstrahlung spectra are also found to be in good agreement with the experimentally observed spectra obtained with various operating anode voltages of the x-ray generator. A computer program has been developed utilizing a few atomic parameters to obtain the elemental concentrations in the sample by a single run using the whole bremsstrahlung. The knowledge of incoming x-ray flux, geometry of experimental arrangements are not required in this technique. To validate the technique, we have taken two samples: an Indian one rupee coin of the year 2000 and a NIST made brass sample, whose compositions are well known. We have used these samples to expose to the bremsstrahlung radiation, produced by the operating anode voltages 20, 25, and 30 KV. The relative concentrations of different elements are determined, which is in good agreement with the earlier results.}
\end{abstract}

\maketitle

\section{Introduction}
Elemental analysis using Energy Dispersive X-ray Fluorescence (EDXRF) technique \citep{mandal2002simple} has several advantages over other techniques like Particle Induced x-ray Emission (PIXE), Neutron Activation Analysis (NAA) etc. A wide range of elements can be detected using EDXRF technique, even one can detect and quantify several elements, simultaneously. Moreover, this is a non-destructive, fast, accurate, and environment friendly technique. To determine the elemental concentration in a specimen with the EDXRF technique, in early days, people used to draw calibration curves \citep{rendle1996r} with several standard reference samples.  With the advancement of computer facilities people started using the alpha-coefficient method \citep{tertian1982principles} where exhaustive regression analysis is required. In this procedure there is a need to know the ratio $(R_i)$ of intensity of a particular element coming from the sample to that from a standard. On the other hand, knowledge of the values of the incident flux, experimental geometry, detector efficiency as well as the photo-electric cross-sections are not required in this method. But for the specimen with a large number of elements, one has to make several standards for each and every element in the sample to determine the $(R_i)$, which is very time consuming as well as costlier. Accurate knowledge of the values of different atomic parameters (e.g. photo-ionisation cross sections, fluorescence yields, Coster-Kronig transitions, radiative widths, etc.) and the availability of high speed computer with large memory, now a days people use fundamental parameter method (FPM) \citep{wilson1959comprehensive} to determine the elemental concentration using EDXRF technique. In this method either one can use mono-chromatic primary radiation coming directly from
a radioactive nucleus (e.g. $^{109}Cd$ and $^{241}Am$), or by using nearly monochromatic fluorescence radiation coming from a secondary target due to the incidence of polychromatic primary radiation, produced by an x-ray generator. In the latter case weighted average energy of the emitted $K_\alpha$ and $K_\beta$ radiation of the secondary target can be treated as mono-energetic radiation which will be considered as the exciting radiation for the sample. Characteristic x-rays coming from different elements present in the sample can be detected using x-ray detector and from the intensities of the characteristic lines one can calculate the elemental concentrations with the help of detector efficiency and few atomic parameters. No standard sample is required in this method. In case of low wattage, small size x-ray generators, emitting x-ray flux is very small and the intensity of the near mono-chromatic secondary radiation is even smaller which in turn will take longer time for accumulating appreciable statistics under different characteristic x-ray lines in the spectrum. So, one can take the whole bremsstrahlung coming from the x-ray generator for exciting the sample instead of taking the fluorescence radiation coming from the secondary target to reduce the experimental time. But due to the presence of a large band energy range in bremsstrahlung it is not an easy task at all. In this paper we would like to demonstrate how one can use the whole bremsstrahlung radiation, coming from a low power, transmission type portable \citep{ferrero2002analysis, gigante1998non} x-ray generator (Amptek), to find the elemental concentrations of the samples using the principles of Fundamental Parameter Method (FPM). As per our knowledge, no attempt has ever been made for EDXRF technique by using ‘white spectrum’\citep{pantenburg1992fundamental} of the bremsstrahlung radiation directly coming from a transmission type x-ray generator.
%%%%%%%%%%%%%%%%%%%%%%%%%%%%%%%%%%%%%%%%%%%%%%%%%%%%%%%%

\section{Experiment}
Here we have used an EDXRF system with a small transmission type portable x-ray generator of Amptek (Mini-X), having maximum operating power of 4 W  with maximum attainable high voltage of 50 KV.
This x-ray generator is made with a metal-ceramic type anode made of silver (Ag) backed with $Al_2O_3$. While doing our experiment, we first record bremsstrahlung spectra directly for three different anode voltages of 20 KV, 25 KV and 30 KV, coming from x-ray tube by placing x-ray detector in front of it at a distance of 12 cm air gap. To clean our required energy region, which is comparatively at the low energy side, we placed two absorbers of tungsten $(W)$ and aluminium $(Al)$ having thicknesses of 1 mil and 10 mil, respectively in front of the x-ray generator. The efficiency of the detector is supplied by the manufacturer. The sample was placed in front of the x-ray generator to irradiate with bremsstrahlung in such a way that the entrance and exit angles of x-rays were fixed at 45$^{\circ}$ with respect to the target plane. X-rays were detected by the Amptek $Si$ pin detector. The resolution of the x-ray detector was 160 eV at 5.9 keV. Two samples that had been used: sample-1: one rupee Indian coin (minted in 2000) and sample-2: a piece of Brass (NIST).

\section{Analysis}

For convenience we split our work into two parts. Firstly, we have fitted the experimental bremsstrahlung spectrum with the appropriate theoretical model for transmission type x-ray tube. Secondly, by using these bremsstrahlung x-rays as an exciter of the sample, we find the constituent elements and their relative concentrations present in the samples.
\subsection{Understanding the Bremsstrahlung Spectrum}
The bremsstrahlung spectrum for thick targets can be found from thin target distribution by calculating electron energy loss with target thickness by collisions with atomic electrons. Target thickness is divided into a large number of thin slices so that each slice contributes to energy loss $\Delta E,$ selected by continuous slowing down approximation assumption.
%In order to obtain bremsstrahlung spectrum for thick target, target thickness is divided in to large number of thin slices so that each slice contributes to energy loss $\Delta E,$ selected by continuous slowing down approximation assumption. 
Thick target bremsstrahlung emission per path length per unit solid angle in the energy interval $k$ to $k+dk$, is
\begin{equation}
\begin{aligned}
I_{E_0},_k=\int_{E>k}^{E_0} I_{E_0},_k,_E(\frac{dE}{-dE/dx})
\label{1}
\end{aligned}
\end{equation}
Where $I_{E_0},_k,_E$ is for thin target bremsstrahlung emission per unit length per unit solid angle, reviewed by Koch and Motz \citep{koch1959bremsstrahlung} in the energy interval $k$ to $k+dk$. $dE/dx$ is the mean electron energy loss per unit path length.
%%%%%%%%%%%%%%%%%%%%%%%%%%%%%%%%%%%%%%%
From equation (\ref{1}), thick target distributions are calculated which is comparable with the Kramers' semi classical non relativistic calculations \citep{storm1972calculated}.
%%%%%%%%%%%%%%%%%%%%%%%%%%%%%%%%%%%%%%%%
%\rrt{After obtaining distributions for thin target, Kramer used Thomas-Widdington law to obtain thick-target energy loss distributions. It shows good agreement with empirical equation of Kulenkampff as follows,  
%\begin{equation}
%\begin{aligned}
%I_{E_0},_k=(27.6/4\pi)\times Z(E_0-k)(ergs/sec-mA-keV-sr)
%\end{aligned}
%\end{equation}
%named as Kramers-Kulenkampff-Dyson (KKD) formula.} 

But electron back scattering losses as well as photon attenuation inside the target are excluded from equation (\ref{1}). 
Electron back-scattering losses is included in equation (\ref{1}) by 

\begin{equation}
\begin{aligned}
I_{E_0},_k=\int_{E>k}^{E_0} I_{E_0},_k,_E(\frac{dE}{-dE/dx})(1-\eta\epsilon_{E0,k,E}) 
\end{aligned}
\end{equation}
Fraction of the incident back-scattered electrons coming from a thick target and total number of electrons averaged over all angles back-scattered with energies between $E$ to $E_0$ are denoted as $\eta$ and $\epsilon_{E_0,k,E}$ respectively. A modified semi-empirical formula of Storm \citep{storm1972calculated}, is used to fit experimental bremsstrahlung is given as, 
\begin{equation}
\begin{aligned}
I_{E_0},_k=\frac{11}{4\pi}\times\frac{Z(E_0-k)(1-e^{-3k/E_k})}{(k/E_0)^\frac{1}{3}(1-e^{-E_0/E_k})}f_{E_0},_k\\\hspace{2cm}[erg/s/mA/keV/sr]
\label{3}
\end{aligned}
\end{equation}
where $I_{E_0},_k$, $Z$, $E_0$ are for bremsstrahlung emission intensity, target atomic number and incident electron energy in keV respectively. $k$ and $E_k$ denote the emitted photon energy and K-shell ionization potential in keV of anode element, respectively. It can be seen that the Kramer-Kulenkampff-Dyson (KKD) model agrees well with experimental values in the range 12-300 keV. 
%Constant can be adjusted in order to match it with theory. 
Here we have used a transmission type x-ray tube for our setup. For which effective penetration length is the width of the effective thickness of the target. To the best of our knowledge no attempt has been made to generate  bremsstrahlung  x-rays theoretically from a transmission type x-ray generator. For transmission type metal-ceramic x-ray tube with silver anode, the photon attenuation correction factor $f_{E_0},{_k}$ is given by

\begin{equation}
    f_{E_0},_k=exp[-\mu_{k}x]
    \label{4}
\end{equation}
where $\mu_{k}$ is the total attenuation coefficient of a photon at energy $k$ in silver anode. Effective thickness of silver anode can be calculated by using the following expression 
\begin{equation}
    x=\frac{{E_0}^2-k^2}{\rho C}
    \label{5}
\end{equation}
\noindent
where the density of silver is denoted by $\rho$. The above-mentioned equations (\ref{4}) and (\ref{5}) are the modified expressions for transmission type x-ray tube. $``C"$ is the Thompson-Whiddington constant which can be extracted by fitting the data given in the paper of Birch and Marshall \citep{birch1979computation} for all the anode voltages.
\newline
To compare the bremsstrahlung emission intensity as given by equation (\ref{3})  with the experimental spectrum, we converted it to the photon number in the energy gap ($\Delta k$) at photon energy $k$. This energy gap is exactly the same as we used during the calibration of our x-ray detector in our present experiment. We maintain the same calibration throughout our experiment for all the operating anode voltages. For conversion of bremsstrahlung emission intensity to photon number \citep{goel1996experimental}, we have used the relation as follows
\begin{equation}
    I_{E_0,k}=\frac{N_k~1.602\times10^3~[erg]}{N_E~\Delta k~(\Omega/4\pi)~\epsilon~[s mA keV sr]}
\end{equation}
\noindent

\noindent
where $N_k$ denotes the total photon number in the energy gap ($\Delta k$). $N_E$, $\Omega$, $\epsilon$ are the total charge (nC) at the target, solid angle  subtended by the x-ray detector at target and the detector efficiency at photon energy k which is supplied by the manufacturer of the detector respectively. 
\newline
Now this photon has to cross the ceramic  ($Al_2O_3$)  part of the metal-ceramic anode. As the thickness of $Al_2O_3$ is not supplied by the manufacturer, we normalised the theoretical spectra with different thicknesses through several trials in order to match the experimental one and used the same thickness of $Al_2O_3$ (nearly 0.25~cm), for all the operating anode voltages. Two absorbers of aluminium and tungsten of thicknesses 1 mil and 10 mil respectively are placed to suppress the low energy side of the spectrum. Moreover, in front of the x-ray tube there is a beryllium window. So, we have to consider the absorption in it also, i.e. in summation $i~=~1$ to $4$ for $Al_2O_3$, $Be$, $Al$ and $W$ in this order respectively. We take all required mass absorption coefficient values from XCOM \citep{berger1987xcom}. Bremsstrahlung emission intensity that emerges out by passing through ceramic part of the anode as well as two absorbers is given by 

\begin{equation}
    N_k^{final}=N_ke^{\large\sum{(-\mu_k)}_i(x)_i} 
\end{equation}

\noindent
%%%%%%%%%%%%%%%%%%%%%%%%%%%%%%%%%%%%%%%%
%\begin{figure}[ht]
%\centering
%\includegraphics[width=8.1cm,height=7cm]{25_EDXRF_ALL.jpg}
%\caption{Experimental (dotted), simulated without putting absorber (orange), simulated (dashed line) without inclusion of detector efficiency which is used in concentration calculation, simulated with inclusion of the efficiency of x-ray detector (red line) bremsstrahlung spectra for 25 KV operating anode voltages of the transmission type x-ray generator.\label{Fig:25_ALL}}
%\end{figure}
%%%%%%%%%%%%%%%%%%%%%%%%%%%%%%%%%%%%%%%%%%
%Fig.~\ref{Fig:25_ALL} shows the raw bremsstrahlung spectrum coming out from x-ray generator for various conditions i.e. before putting the absorbers, after putting absorbers but without taking detector efficiency contribution, after taking detector efficiency contribution only for 25 KV anode voltage. 
%%%%%%%%%%%%%%%%%%%%%%%%%%%%%%%%%%%%%%%%
\begin{figure}[ht]
\centering
\includegraphics[width=8.1cm,height=8.1cm]{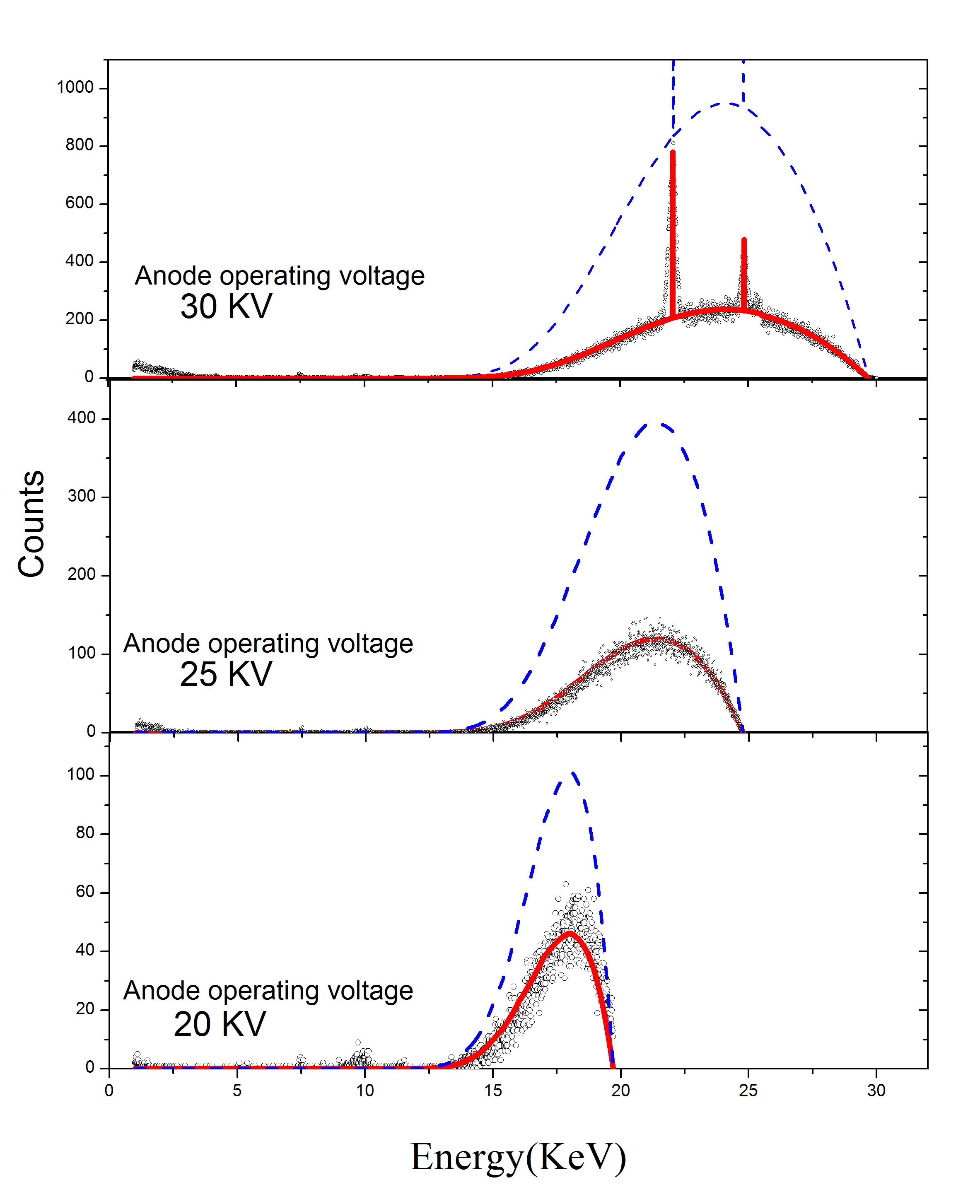}
\caption{Experimental (dotted), theoretically generated without taking detector efficiency (dashed line) and with the inclusion of the efficiency of x-ray detector (red line) bremsstrahlung spectra for various operating anode voltages of the transmission type x-ray generator.\label{Fig:diamond}}
\end{figure}
%%%%%%%%%%%%%%%%%%%%%%%%%%%%%%%%%%%%%%%%%%
Fig.~\ref{Fig:diamond} shows the bremsstrahlung spectrum coming out from x-ray generator without taking detector efficiency contribution, after taking detector efficiency contribution for three different operating anode voltages (20 KV, 25 KV, 30 KV). We have also generated the characteristic lines of silver $K$ x-rays due to the silver anode with the help of reference \citep{ebel1999x, pella1985analytical}.
However, contributions due to the characteristic lines, used as an exciter, is negligible in comparison to the continuous bremsstrahlung spectra as the area under the characteristic lines are very small with respect to the continuous one. We have done this experiment for various operating anode voltages of 20 KV, 25 KV, 30 KV and for each case the experimental spectrum is perfectly matched with the theoretical one. As the $K$-shell binding energy of Silver (Ag) is 25.514 keV, so among our three operating voltages we can see the characteristic lines for only the tube voltage of 30 KV.  
\subsection{Concentration Calculation}

Now while considering the elemental analysis by mono-chromatic x-rays, fluorescent x-ray line intensity of an element ($I_i$) \citep{birch1979computation} can be written as

\begin{eqnarray}
     I_i&=&\frac{I_0\Omega}{4\pi sin  \theta_1}~[\sigma_i\omega_if_i]~A_i\epsilon_iC_i~(1+H_i)\nonumber\\
     &=&S~[\sigma_i\omega_if_i]~A_i\epsilon_iC_i~(1+H_i)\nonumber\\
     &=&SV_i C_i
     \label{8}
\end{eqnarray}
%%%%%%%%%%%%%%%%%%%%%%%%%%%%%%%%%%%%%%%%%%%%
\begin{figure*}[ht]
\centering
\includegraphics[width=8.1cm,height=8.1cm]{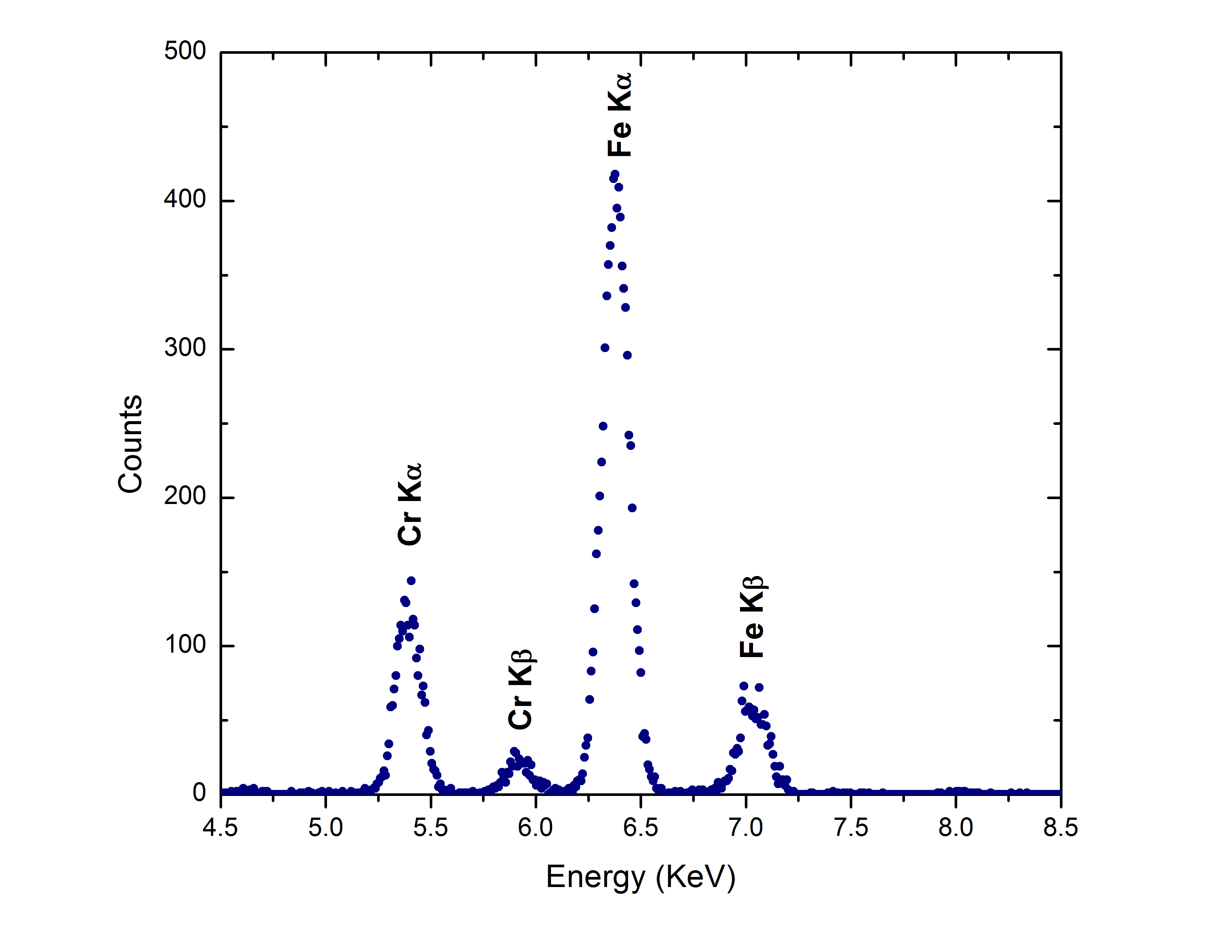}
\includegraphics[width=8.1cm,height=8.1cm]{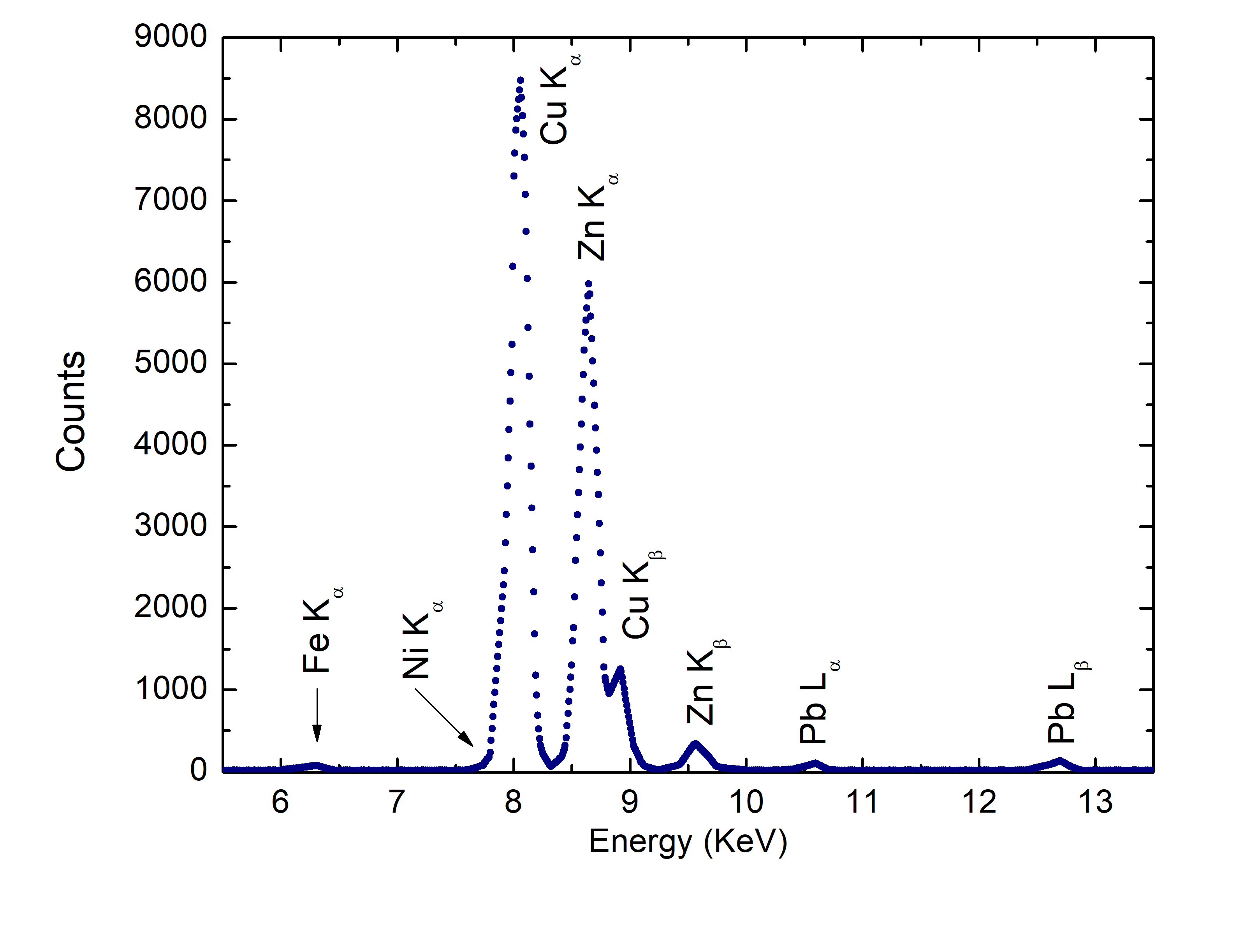}
\caption{A typical spectrum obtained due to the bremsstrahlung x-ray irradiation, from a   transmission type x-ray generator operating at 25 KV anode voltage, of an Indian one rupee coin [2000](left). A typical spectrum of Brass Sample [NIST] obtained due to the bremsstrahlung x-ray irradiation, from a transmission type x-ray generator operating at 25 KV operating anode voltage(right).\label{Fig:Sample}}
\end{figure*}
%%%%%%%%%%%%%%%%%%%%%%%%%%%%%%%%%%%%%%%%%%
\noindent Where $I_0$, $\Omega$, $\theta_1$ are for incident flux, solid angle subtended by the detector with respect to the target and entrance angle, respectively. $\sigma_i$ and $\omega_i$ denote photo-ionization cross section of the $i^{th}$ element for primary radiation and fluorescence yield. $f_i$ is the fraction of radiation of the considered x-ray line and $A_i$ is the absorption correction not only for the primary x-rays but also for the fluorescent x-rays inside the target. $\epsilon_i$ is the detector efficiency for fluorescent radiation, $H_i$ is the inter-element enhancement factor and $C_i$ is for relative concentration for $i^{th}$ element of the sample. We calculate $I_i$ by using the area under the $K_\alpha$ line of each sample. The other terms can be explained as 
\begin{eqnarray}
    S&=&\frac{I_0\Omega}{4\pi sin\theta_1} \label{9}\\
    A_i&=&\frac{1-exp(-[\mu_s^0/sin \theta_1 + \mu_s^i/sin \theta_2]t)}{[\mu_s^0/sin \theta_1 +  \mu_s^i/sin \theta_2]}\\
     \mu_s^0&=&\Sigma C_i\mu^0;\\
     \mu_s^i&=&\Sigma C_i\mu^i;\\
   V_i&=&[\sigma_i\omega_if_i]A_i\epsilon_i(1+H_i)\noindent
\end{eqnarray}
where $\theta_2$ is for exit angle, $``t"$ denotes the thickness of the sample, $\mu_s^0$ and $\mu_s^i$ denote the mass absorption coefficients of the sample for primary and fluorescent radiations, respectively. Inter-element enhancement factor $H_i$ for infinitely thick sample can be easily calculated by analytically developed expression of \textcite{sparks1975quantitative} as given by 
\begin{equation}
\begin{aligned}
    H_i = \frac{{C_j} {{[\sigma_0]_j[ \omega_k f_k]}_j} {[\sigma_j]}_i} {2 {[\sigma_0]}_i}\times\bigg\{ [sin\theta_1/{{\mu_s}^0}]              ln\bigg(\frac{\frac{{{\mu_s}^0}}{sin{\theta_1}}}{{\mu_s}^j}+1\bigg)\\ + [sin\theta_2/{{\mu_s}^0}]              ln\bigg(\frac{\frac{{{\mu_s}^0}}{sin{\theta_2}}}{{\mu_s}^j}+1\bigg)\bigg\}
\end{aligned}
\end{equation}
\noindent
This enhancement factor denotes the effect of excitation of the $i^{th}$ element due to the $j^{th}$ element. $[\sigma_j]_i$ is for photo-ionization cross-section of the $i^{th}$ element due to the $j^{th}$ element whereas $[\sigma_0]_i$ and $[\sigma_0]_j$ are the same for $i^{th}$ and $j^{th}$ element due to the primary radiation. This factor depends on the matrix element inside the target. For our target this factor contributes a very little amount, much less than $1\%$ to the concentration calculation. This contribution is so small that we neglect the contribution coming from all other characteristic lines of element $``j"$ that can excite atoms of element $``i"$ in the relevant shell. 
%%%%%%%%%%%%%%%%%%%%%%%%%%%%%%%%%%%%%%%
\begin{table}[]
\resizebox{8.3cm}{!}{
\begin{tabular}{|l|l|l|l|}
\hline
\multicolumn{4}{|l|}{\multirow{2}{*}{\begin{tabular}[c]{@{}l@{}}\hspace{0.38 cm}Results of an Indian one rupee coin(2000) made of\\                        \hspace{3.2 cm}Cr and Fe\end{tabular}}} \\
\multicolumn{4}{|l|}{} \\ \hline
\begin{tabular}[c]{@{}l@{}}Anode \\ Voltage\\   (in KV)\end{tabular} & \begin{tabular}[c]{@{}l@{}}\hspace{0.7cm}Our \\        Results (in wt\%)\end{tabular} & \begin{tabular}[c]{@{}l@{}}\hspace{0.4cm}Average \\     Value (in wt\%)\end{tabular} & \begin{tabular}[c]{@{}l@{}}Earlier Results       \\       (in wt\%)   {[}22{]}\end{tabular} \\ \hline
\hspace{0.36 cm}20 & \begin{tabular}[c]{@{}l@{}}Cr 17.1 $\pm$ 1.5\\ Fe 82.9$\pm$ 4.2\end{tabular} & \multirow{3}{*}{\begin{tabular}[c]{@{}l@{}}17.1 $\pm$ 1.4\\ 82.9 $\pm$ 4.2\end{tabular}} & \multirow{3}{*}{\begin{tabular}[c]{@{}l@{}}Cr 16.8 $\pm$ 1.2\\ Fe 83.2 $\pm$ 5.8\end{tabular}} \\ \cline{1-2}
\hspace{0.36 cm}25 & \begin{tabular}[c]{@{}l@{}}Cr 17.2 $\pm$ 1.4\\ Fe 82.8 $\pm$ 4.1\end{tabular} &  &  \\ \cline{1-2}
\hspace{0.36 cm}30 & \begin{tabular}[c]{@{}l@{}}Cr 16.9 $\pm$1.4\\ Fe 83.1 $\pm$ 4.3\end{tabular} &  &  \\ \hline
\end{tabular}}
\caption{Concentrations of the compositions of an Indian one rupee coin (minted in 2000)\label{Table:t1}}
\end{table}
%%%%%%%%%%%%%%%%%%%%%%%%%%%%%%%%%%%%%%%%%%%%%%
\begin{table}[]
\resizebox{8.3cm}{!}{
\begin{tabular}{|l|l|l|l|}
\hline
\multicolumn{4}{|l|}{\hspace{2.4 cm}Results of Brass Sample (NIST)}                                                  \\ \hline
\begin{tabular}[c]{@{}l@{}}Anode Voltage\\        \hspace{0.5 cm}(KV)\end{tabular} & \begin{tabular}[c]{@{}l@{}}\hspace{0.3 cm}Our Results\\ \hspace{0.38 cm}(in wt \%)\end{tabular}                                                                                              & \begin{tabular}[c]{@{}l@{}}Average Value\\ \hspace{0.38 cm} (in wt\%)\end{tabular}                                                                                                             & \begin{tabular}[c]{@{}l@{}}\hspace{0.18 cm}Given\\ (in wt \%)\end{tabular}                                                                  \\ \hline
\hspace{0.75 cm}20                                                                  & \begin{tabular}[c]{@{}l@{}} Fe 0.20 $\pm$ 0.02\\ Ni 0.10 $\pm$ 0.02\\ Cu 62.5 $\pm$ 4.3\\ Zn 35.3 $\pm$ 4.3\\ Sn  Not detected\\ Pb 1.90 $\pm$ 0.2 \end{tabular} & \multirow{3}{*}{\begin{tabular}[c]{@{}l@{}} Fe  0.20 $\pm$ 0.02\\ Ni  0.10 $\pm$ 0.02\\ Cu 62.8 $\pm$ 4.3\\ Zn 35.1 $\pm$ 4.3\\ Sn   Not detected\\ Pb 1.8 $\pm$ 0.2\end{tabular}} & \multirow{3}{*}{\begin{tabular}[c]{@{}l@{}}Fe  0.088\\ Ni  0.07\\ Cu  61.33\\ Zn  35.31\\ Sn  0.43\\ Pb  2.77\end{tabular}} \\ \cline{1-2}
\hspace{0.75 cm}25                                                                  & \begin{tabular}[c]{@{}l@{}}Fe 0.20 $\pm$ 0.02\\ Ni 0.10 $\pm$ 0.02\\ Cu 62.9 $\pm$ 4.3\\ Zn 35.1 $\pm$ 4.3\\ Sn  Not detected\\ Pb 1.7 $\pm$ 0.2\end{tabular}  &                                                                                                                                                                                   &                                                                                                                             \\ \cline{1-2}
\hspace{0.75 cm}30                                                                  & \begin{tabular}[c]{@{}l@{}}Fe 0.20 $\pm$ 0.02\\ Ni 0.10 $\pm$ 0.02\\ Cu 63.1 $\pm$ 4.3\\ Zn 34.8 $\pm$ 4.3\\ Sn  Not detected\\ Pb 1.8 $\pm$ 0.2\end{tabular}  &                                                                                                                                                                                   &                                                                                                                             \\ \hline
\end{tabular}}
\caption{Concentrations of the compositions of Brass Sample [NIST]\label{Table:t2}}
\end{table}
%%%%%%%%%%%%%%%%%%%%%%%%%%%%%%%%%%%%%%
Now from equation (\ref{8}) we can write 
\begin{equation}
    C_i=\frac{I_i}{SV_i}
    \label{15}
\end{equation}
We have assumed that all the detectable elements will add up to 100\%. To fulfill this condition we must use 
\begin{equation}
   \sum C_i=1
   \label{16}
\end{equation}
So, by using the set of equations (\ref{9}) to (\ref{16}), assuming an initial arbitrary concentration of each element we can find the corresponding modified concentrations, which is used as assumed concentrations for the next iteration. After each iteration, concentration of each elements $(C_i)$ is normalized by 
\begin{equation}
    C_i=\frac{C_i{^\prime}}{\sum C_i{^\prime}}
\end{equation}
The above mentioned procedure is applicable only when mono-energetic x-rays are used as an exciter. But here we used bremsstrahlung coming from a transmission type x-ray tube to irradiate the sample. In order to use the above mentioned procedure we divide the whole bremsstrahlung spectrum into a large number of energy slices so that each energy slice can be treated as a source of mono-energetic x-ray beam. The width of this energy slice is the same as the energy per unit channel in our present experiment. We have calculated the concentrations ($C_i$) for all the energy slices separately and add them up and normalize to get the modified one for the next iteration if needed.
\section{Results and discussions}

\noindent
Fig.~\ref{Fig:Sample} show the typical x-ray spectrum of a coin and brass sample which we got due to the irradiation of those samples by the bremsstrahlung radiation coming directly from a transmission type x ray tube operated at 25 KV anode voltage.

\noindent
By using the above procedure the concentrations of each elements in that two used samples have been calculated and the results are shown in Table.~\ref{Table:t1} and Table.~\ref{Table:t2} along with the previously published results \citep{mandal2003edxrf} and NIST Brass Sample compositions in which they used monochromatic x-rays as incident beam. Both the results agree well within the error bar.
\noindent From equation \ref{15}, it seems that the error in the concentration comes from only the terms $I_i$ and $V_i$. The term $S$ cancels out while normalizing the concentrations.
Errors are almost negligible for the elements with higher concentrations, and written as follows,
%%%%%%%%%%%%%%%%%%%%%%%%%%%%%%%
\begin{equation}
    (\frac{\Delta C_i}{C_i})=\sqrt{(\frac{\Delta I_i}{I_i})^2+(\frac{\Delta V_i}{V_i})^2}
    \label{18}
\end{equation}
%%%%%%%%%%%%%%%%%%%%%%%%%%%%%%%%%
The error in $V_i$ is propagated through the errors in (i) photo-ionization cross section of k-shell ($\sim$2\%) and (ii) atomic parameters ($\omega, f$) of k-shell ($\sim$3\%), (iii) mass attenuation coefficients ($\sim$2\%). 
%%%%%%%%%%%%%%%%%%%%%%%%%%%%%%%
\begin{equation}
    (\frac{\Delta V_i}{V_i})=\sqrt{(\frac{\Delta \sigma_i}{\sigma_i})^2+(\frac{\Delta \omega_i}{\omega_i})^2+(\frac{\Delta f_i}{f_i})^2+(\frac{\Delta \epsilon_i}{\epsilon_i})^2+(\frac{\Delta A_i}{A_i})^2}
    \label{19}
\end{equation}
%%%%%%%%%%%%%%%%%%%%%%%%%%%%%%%%
Neglecting the error in $H_i$ due to its small contribution, the error in $C_i$ turns out to be within 3-8\%. Overall errors in the concentration for each element is listed in table \ref{Table:t1} and \ref{Table:t2}.\\
\noindent
Thus it can be seen that one can easily find the concentrations of each element of the sample by irradiating them with bremsstrahlung radiation directly coming from a portable transmission type x-ray tube without using any secondary source as well as any standard references.
\newline
The main advantages of this type of setup are (i) a very low powered portable x-ray tube can be used, (ii) it can be operated by using a stable power source like a car battery, and (iii) with a laptop.

\section{Acknowledgements}
The authors would like to acknowledge DST-FIST for funding the procurement of the instruments and University of Kalyani for providing necessary infrastructural facility.

%%%%%%%%%%%%%%%%%%%%%%%%%%

\bibliographystyle{unsrtnat}
\bibliography{main.bbl}
\end{document}